\documentclass[prb,10pt,twocolumn]{revtex4}

\usepackage{epsfig}
\graphicspath{{/}}

\newcommand{\fracd}[2]{\frac{\displaystyle{#1}}{\displaystyle{#2}}}

\begin{document}

\title{Ultrafast Carrier Recombination and Generation Rates for Plasmon Emission and Absorption in Graphene}

\author{Farhan Rana, Jared H. Strait, Haining Wang, Christina Manolatou}
\affiliation{School of Electrical and Computer Engineering, Cornell University, Ithaca, NY 14853}
\email{fr37@cornell.edu}

\begin{abstract}
Electron-hole generation and recombination rates for plasmon emission and absorption in graphene are presented. The recombination times of carriers due to plasmon emission have been found to be in the tens of femtoseconds to hundreds of picoseconds range. The recombination times depend sensitively on the carrier energy, carrier density, temperature, and the plasmon dispersion. Carriers near the Dirac point are found to have much longer lifetimes compared to carriers at higher energies. Plasmons in a graphene layer on a polar substrate hybridize with the surface optical phonons and this hybridization modifies the plasmon dispersion. We also present generation and recombination rates of carriers due to plasmon emission and absorption in graphene layers on polar substrates. 
\end{abstract}

\maketitle

\section{Introduction}
The high carrier mobility and the large optical absorption in graphene have opened up unique opportunities for this material in electronics and optoelectronics~\cite{nov1,nov2,dressel1,heer,shepard,marcus,rana,ryzhii,avouris,ryzhii2,ryzhii3,Bon,avouris2,ryzhii4,ryzhii5,ryzhii6}. The performance of graphene in many of these applications depends on the electron-hole generation and recombination rates in graphene. It is therefore important to understand the mechanisms that are responsible for electron-hole generation and recombination in graphene and the associated time scales. Previously, carrier generation and recombination rates in graphene due to Auger scattering and impact ionization and due to optical phonon emission and absorption have been reported~\cite{rana2, rana3}. The strong interaction between electrons/holes and plasmons in graphene has been used to explain observed features in the angled resolved photoemission (ARPES) data~\cite{bostwick, ohta, polini, sarma0}. In this paper, we present electron-hole generation and recombination rates due to plasmon emission and absorption. Our results show that the recombination times of carriers due to plasmon emission are in the tens of femtoseconds to hundreds of picoseconds range and depend sensitively on the carrier energy, carrier density, temperature, and the plasmon dispersion. The available phase space for plasmon emission is restricted because of energy and momentum conservation requirements and also because of Pauli's exclusion principle and carriers near the Dirac point have much longer lifetimes compared to carriers at higher energies. Plasmons in a graphene layer on a polar substrate (Figure (\ref{fig0}b)) hybridize with the surface optical phonons and this hybridization splits the plasmon dispersion into two branches~\cite{koch,sarma}. We also present generation and recombination rates of carriers due to plasmon emission and absorption in graphene layers on polar substrates. The results presented here, in the light of the previous studies~\cite{rana2,rana3}, indicate that plasmon emission is the dominant mechanism for carrier recombination in graphene. Our results are expected to be useful in interpreting experimental observations in ultrafast optical studies~\cite{rana4,rana5,heinz,heinz2,norris,driel,rana6,kaindl} and in understanding the operation of graphene based optoelectronic devices~\cite{avouris,avouris2,mceuen,park}.

\section{Theoretical Model for Suspended Graphene}
\begin{figure}[tbp]
  \begin{center}
   \epsfig{file=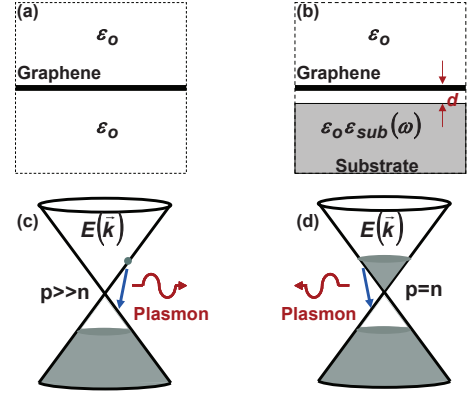,angle=0,width=2.5 in}    
    \caption{(a) A suspended graphene sheet. (b) A graphene sheet on a polar substrate. (c) Electron-hole recombination via plasmon emission in p-doped graphene ($p>>n$). (d) Electron-hole recombination via plasmon emission in photoexcited graphene ($n=p$).}
    \label{fig0}
  \end{center}
\end{figure}

We first consider a graphene sheet in the plane $z=0$ sandwiched by media with free-space permittivity (Figure (\ref{fig0}a)), as in the case of suspended graphene~\cite{bolotin}  . The electron energy dispersion are given by $E_{s}(\vec{k})= s \hbar v k$, where $s$ equals +1 and -1 for conduction and valence bands, respectively. The dispersion for the plasmons in given by $\epsilon(q,\omega)=0$, where~\cite{rana,sarma2,tapash},
\begin{equation}
\epsilon(q,\omega) = 1 - \fracd{e^{2}}{2 \epsilon_{o} q} \Pi(q,\omega) \label{eq:epPi}
\end{equation}
The electron-hole propagator $\Pi(q,\omega)$ is~\cite{rana,sarma2,tapash},
\begin{eqnarray}
\Pi(q,\omega) & = & 4  \sum_{s,s'} \int \fracd{d^{2}\vec{k}}{(2\pi)^{2}} \, |\langle \psi^{s'}_{\vec{k}+\vec{q}}(\vec{r}) | e^{i\vec{q}.{\vec{r}}} | \psi^{s}_{\vec{k}}(\vec{r}) \rangle|^{2} \nonumber \\
& \times & \fracd{f_{s}(\vec{k}) -  f_{s'}(\vec{k}+\vec{q})}{\hbar \omega + E_{s}(\vec{k}) - E_{s'}(\vec{k} +\vec{q}) + i\eta}  \label{eq:Pi}
\end{eqnarray}
The matrix element between the Bloch functions in the above expression equals~\cite{rana,sarma2,tapash}, 
\begin{equation}
|\langle \psi^{s'}_{\vec{k}+\vec{q}}(\vec{r}) | e^{i\vec{q}.{\vec{r}}} | \psi^{s}_{\vec{k}}(\vec{r}) \rangle|^{2} = \fracd{1}{2} \left[ 1 + ss'\fracd{k+q\cos(\theta)}{|\vec{k}+\vec{q}|} \right]
\end{equation}
Here, $\theta$ is the angle between $\vec{k}$ and $\vec{q}$. To calculate the recombination and generation rates, we consider a plasmon wave with the electric field given by,
\begin{equation}
\vec{E}(\vec{r},z,t) = \fracd{1}{2} (\hat{q} \pm i \hat{z}) E_{o} e^{\mp|\vec{q}|z} e^{i\vec{q}.\vec{r} - i\omega(q) t} + c.c.
\end{equation}
The transition rate for an electron in the conduction band to go into the valence band via stimulated emission of a plasmon of wavevector $\vec{q}$ is given by the Fermi's Golden Rule,
\begin{eqnarray}
& \fracd{1}{\displaystyle \tau_{\vec{k}}} = \fracd{2\pi}{\displaystyle \hbar} |\langle \psi^{-}_{\vec{k}-\vec{q}}(\vec{r}) | e^{-i\vec{q}.{\vec{r}}} | \psi^{+}_{\vec{k}}(\vec{r}) \rangle|^{2} \fracd{e^{2} |E_{o}|^{2}}{4 q^{2}}  & \nonumber \\
& \times  (1 - f_{-}(\vec{k}-\vec{q})) \delta(E_{+}(\vec{k}) - E_{-}(\vec{k}-\vec{q}) - \hbar \omega(q)) & \label{eq:time1}
\end{eqnarray}
The energy density $W$ of the plasmon wave has contributions from both the field as well as the kinetic energy of the carriers. Assuming no plasmon dissipation, the total energy density can be found from the complex electromagnetic energy theorem~\cite{kong},
\begin{eqnarray}
W & = & W_{F} + W_{KE} \nonumber \\
& = & \fracd{\epsilon_{o}}{2q} |E_{o}|^{2} -  \fracd{1}{4} |E_{o}|^{2} \, \Im \left\{ \left. \fracd{\partial \sigma(q,\omega)}{\partial \omega} \right|_{\omega(q)} \right\} 
\end{eqnarray}
Since the conductivity $\sigma(q,\omega)$ is related to the dielectric constant $\epsilon(q,\omega)$ as,
\begin{equation}
\epsilon(q,\omega) = 1 + i \fracd{q \sigma(q,\omega) }{2\epsilon_{o} \omega}
\end{equation}
the expression for the energy density $W$ becomes,
\begin{equation}
W = \fracd{\epsilon_{o}}{2q} |E_{o}|^{2} \Re \left\{ \left. \omega \fracd{\partial \epsilon(q,\omega)}{\partial \omega} \right|_{\omega(q)} \right\} \label{eq:W}
\end{equation}
$W$ must also equal $n(\vec{q}) \hbar \omega(q)/A$, where $n(\vec{q})$ is the number of plasmons in the mode $\vec{q}$ and $A$ is the area of the crystal. Therefore, using (\ref{eq:time1}) and (\ref{eq:W}), the lifetime of an electron in the conduction band due to both stimulated and spontaneous emission into all plasmon modes becomes,
\begin{eqnarray}
\fracd{1}{\displaystyle \tau_{\vec{k}}} & = & \fracd{2\pi}{\displaystyle \hbar} \int \fracd{d^{2}\vec{q}}{(2\pi)^{2}} (n(\vec{q})+1) (1 - f_{-}(\vec{k}-\vec{q})) \nonumber \\
& \times & \fracd{e^{2}}{2\epsilon_{o} q} \fracd{1}{2} \left[ 1 - \fracd{k-q\cos(\theta)}{|\vec{k}-\vec{q}|} \right]  \nonumber \\
& \times & \fracd{\hbar \delta(E_{+}(\vec{k}) - E_{-}(\vec{k}-\vec{q}) - \hbar \omega(q))}{  \Re \left\{ \left. \fracd{\partial \epsilon(q,\omega)}{\partial \omega} \right|_{\omega(q)} \right\}  }  \label{eq:time2}
\end{eqnarray}
The recombination and generation rates, $R$ and $G$ (units: $\#$/cm$^{2}$-s) due to plasmon emission and absorption can be written as,
\begin{eqnarray}
R & = &  8 \pi \int \fracd{d^{2}\vec{k}}{(2\pi)^{2}} \int \fracd{d^{2}\vec{q}}{(2\pi)^{2}} (n(\vec{q})+1)  \nonumber \\
& \times & f_{+}(\vec{k}) (1 - f_{-}(\vec{k}-\vec{q})) \nonumber \\ 
& \times & \fracd{e^{2}}{2\epsilon_{o} q} \fracd{1}{2} \left[ 1 - \fracd{k-q\cos(\theta)}{|\vec{k}-\vec{q}|} \right]  \nonumber \\
& \times & \fracd{\delta(E_{+}(\vec{k}) - E_{-}(\vec{k}-\vec{q}) - \hbar \omega(q))}{  \Re \left\{ \left. \fracd{\partial \epsilon(q,\omega)}{\partial \omega} \right|_{\omega(q)} \right\}  }  \label{eq:R}
\end{eqnarray}
\begin{eqnarray}
G & = &  8 \pi \int \fracd{d^{2}\vec{k}}{(2\pi)^{2}} \int \fracd{d^{2}\vec{q}}{(2\pi)^{2}} n(\vec{q})  \nonumber \\
& \times & (1-f_{+}(\vec{k})) f_{-}(\vec{k}-\vec{q}) \nonumber \\ 
& \times & \fracd{e^{2}}{2\epsilon_{o} q} \fracd{1}{2} \left[ 1 - \fracd{k-q\cos(\theta)}{|\vec{k}-\vec{q}|} \right]  \nonumber \\
& \times & \fracd{\delta(E_{+}(\vec{k}) - E_{-}(\vec{k}-\vec{q}) - \hbar \omega(q))}{  \Re \left\{ \left. \fracd{\partial \epsilon(q,\omega)}{\partial \omega} \right|_{\omega(q)} \right\}  }  \label{eq:G}
\end{eqnarray}
In thermal equilibrium, the plasmon number $n(\vec{q})$ equals the Bose factor $(\exp(\hbar \omega(q)/KT)-1)^{-1}$. Similar results can be obtained starting from the self-energy of an electron in the conduction band. Assuming thermal equilibrium, and using the imaginary-time Green's function approach, the relevant contribution to the conduction band electron self-energy can be written as~\cite{mahan},
\begin{eqnarray}
{\textstyle \sum} (\vec{k},i\omega_{n}) =  -  \int \fracd{d^{2}\vec{q}}{(2\pi)^{2}}  \fracd{1}{2} \left[ 1 - \fracd{k-q\cos(\theta)}{|\vec{k}-\vec{q}|} \right] \, \,\,\,\,\,\, \nonumber & & \\
 \times \frac{1}{\beta \hbar} \sum_{m} \fracd{e^{2}/2\epsilon_{o} q}{\epsilon(q,i\nu_{m})} G(\vec{k}-\vec{q},i\omega_{n} - i\nu_{m}) & & \label{eq:self}
\end{eqnarray}
where $G(\vec{k}-\vec{q},\omega_{n} - i\nu_{m})$ is the valence band Green's function. The summation over the Matsubara frequencies can be performed by first isolating the pole coming from the zero of $\epsilon(q,i\nu_{m})$ in the denominator at the plasmon frequency. Finally, if one excludes from (\ref{eq:self}) contributions coming from plasmon absorption processes and then calculates the lifetime of the conduction electron using the expression~\cite{mahan},
\begin{equation}
\fracd{1}{\displaystyle \tau_{\vec{k}}} = -\frac{2}{\hbar} \Im \left\{ {\textstyle \sum} (\vec{k},(E_{+}(\vec{k})-E_{f})/\hbar + i\eta) \right\} \label{eq:time3}
\end{equation}
then the result obtained is identical to the one given earlier in (\ref{eq:time2}). It should be mentioned here that focusing on the collective excitation pole coming from the zero of $\epsilon(q,i\nu_{m})$ allows one to calculate the interband scattering rate due to electron-plasmon interaction. Contributions from other processes, such as Auger scattering and impact ionization~\cite{rana2}, are therefore excluded. From the results obtained above it follows that the electron-plasmon interaction in graphene can be approximately described by the following Hamiltonian in the second quantized form,
\begin{equation}
\hat{H}_{el-pl} = \sum_{s,s',\sigma,\vec{k},\vec{q}} M_{s,s',\vec{k},\vec{q}} \left( \hat{b}_{\vec{q}} +  \hat{b}^{\dagger}_{-\vec{q}} \right) \, \hat{c}^{\dagger}_{s', \sigma, \vec{k}+\vec{q}} \, \hat{c}_{s, \sigma, \vec{k}} 
\end{equation}
where, $\hat{b}_{\vec{q}}$ and $\hat{c}_{s,\sigma,\vec{k}}$ are the plasmon and the electron destruction operators, respectively, $\sigma$ stands for different spins and valleys, and the coupling constant $M_{s,s',\vec{k},\vec{q}}$ is given by,
\begin{equation}
\left| M_{s,s',\vec{k},\vec{q}} \right|^{2} = \fracd{e^{2} \, \hbar}{2\epsilon_{o} \,q \, A} \fracd{   \fracd{1}{2} \left[ 1 + ss'\fracd{k+q\cos(\theta)}{|\vec{k}+\vec{q}|} \right]       }{  \Re \left\{ \left. \fracd{\partial \epsilon(q,\omega)}{\partial \omega} \right|_{\omega(q)} \right\} }
\end{equation}
Here, $\theta$ is the angle between $\vec{k}$ and $\vec{q}$ and $A$ is the area of the graphene crystal.   
\begin{figure}[t]
  \begin{center}
   \epsfig{file=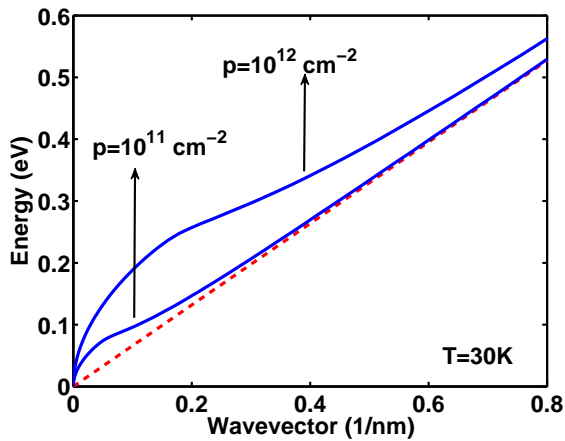,angle=-0,width=3.0 in}    
    \caption{The plasmon dispersion (solid) in a p-doped graphene is shown for different hole densities. The dashed curve represents $\hbar v q$. $T=30K$.}
    \label{fig1}
  \end{center}
\end{figure}
\begin{figure}[t]
  \begin{center}
   \epsfig{file=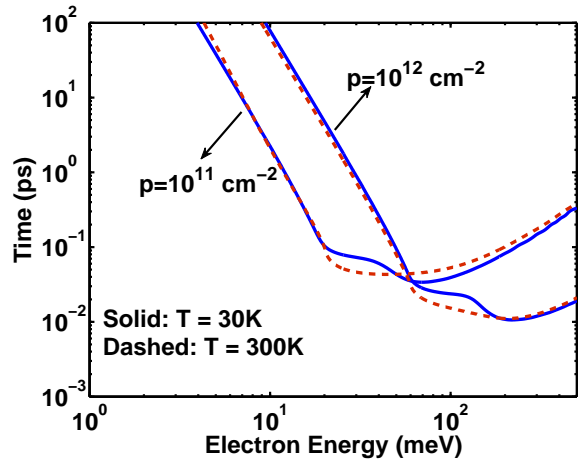,angle=-0,width=3.0 in}    
    \caption{The calculated spontaneous emission lifetime of an electron in the conduction band is plotted as a function of the electron energy for different hole densities and temperatures in p-doped graphene.}
    \label{fig2}
  \end{center}
\end{figure}

\subsection{Results and Discussion}
The plasmon dispersion is first found numerically using the expression for $\Pi(q,\omega)$ in (\ref{eq:Pi}). The recombination and generation rates and lifetimes are then calculated using (\ref{eq:time2}), (\ref{eq:R}), and (\ref{eq:G}). The dominant contribution to the propagator in (\ref{eq:Pi}) comes from the intraband part. The interband part modifies the plasmon dispersion slightly and also imparts an imaginary part to the plasmon frequency.  If the interband contribution is ignored the error in the calculated plasmon frequency has been found to be generally small (less than 10$\%$) for small plasmon wavevectors. This small error comes with the enormous simplicity of having to find zeroes of $\epsilon(q,\omega)$ on only the real frequency axis and therefore this approach has been adopted in numerical simulations. The results we present are also not self-consistent in the sense that the quasiparticle density of states have been assumed to be that of the non-interacting electron system. It is known that electron-plasmon interaction can modify the quasiparticle density of states and generate plasmaron bands~\cite{bostwick}. However, the modification of the quasiparticle density of states is expected to be small for the electron and hole densities considered in this paper. The average recombination and generation times are defined as, $\tau_{R}^{-1} = R/{\rm min} (n, p)$, and $\tau_{G}^{-1} = G/{\rm min} (n, p)$, respectively, where $n$ and $p$ are the electron and hole densities. 
\begin{figure}[bp]
  \begin{center}
   \epsfig{file=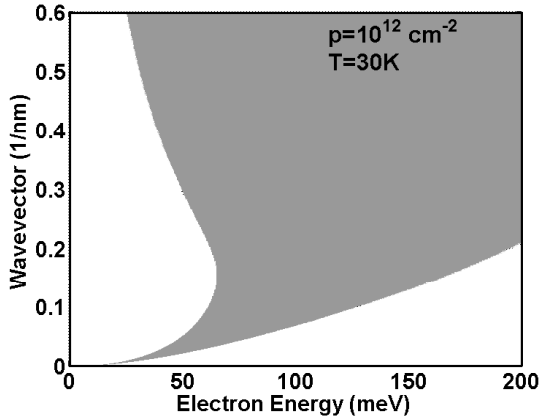,angle=-0,width=3.0 in}    
    \caption{The shaded region shows the allowed values of the spontaneously emitted plasmon wavevectors as a function of the electron energy in the conduction band for p-doped graphene. $p=10^{12}$ cm$^{-2}$ and $T=30K$}
    \label{fig3}
  \end{center}
\end{figure}
\begin{figure}[tbp]
  \begin{center}
   \epsfig{file=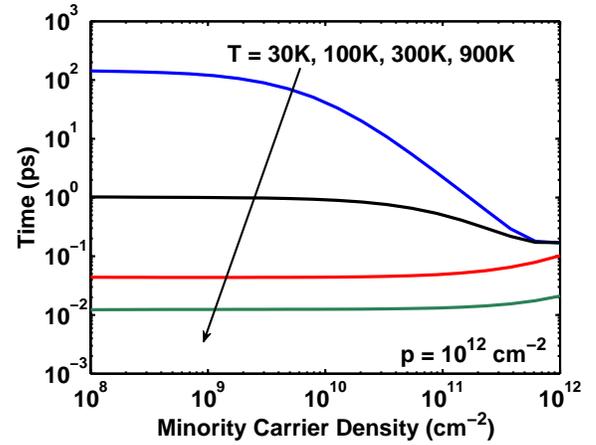,angle=-0,width=3.0 in}    
    \caption{The average minority carrier (electron) recombination time $\tau_{R}$ is plotted as a function of the minority carrier density for different temperatures in p-doped graphene ($p=10^{12}$ cm$^{-2}$). The arrow indicates curves for increasing values of the temperature ($T=30,100,300,900$K).}  
    \label{fig4}
  \end{center}
\end{figure}
Figure (\ref{fig1}) shows the plasmon dispersion in p-doped graphene for different hole densities. Figure (\ref{fig2}) shows the calculated lifetime of an electron in the conduction band due to spontaneous plasmon emission as a function of the electron energy for different hole densities and temperatures in p-doped graphene (Figure (\ref{fig0}c)). Note that the electron-hole symmetry in graphene implies that the hole lifetimes in n-doped graphene would be identical. Figure (\ref{fig2}) shows that conduction electrons with energies near the Dirac point have much longer lifetimes compared to the electrons at higher energies. This trend can be understood as follows. Energy and momentum conservation require that an electron with wavevector $\vec{k}$ can emit a plasmon with wavevector $\vec{q}$ only if,
\begin{equation}
\fracd{1}{2}\left(\fracd{\omega(q)}{v} - q \right) \le k \le \fracd{1}{2}\left(\fracd{\omega(q)}{v} + q \right)
\end{equation}
The shaded region in Figure (\ref{fig3}) shows the allowed values of the wavevector of the emitted plasmon as a function of the electron energy assuming $p=10^{12}$ cm$^{-2}$ and $T=30K$. As the electron energy becomes smaller than $\sim$60 meV, the allowed phase space for the plasmon wavevectors shrinks significantly for small wavevectors. The probabilities of emission of plasmons of large wavevectors are small because the energies of such plasmons are large and the resulting final electron states deep inside the valence band are already occupied by valence electrons. In addition, numerical simulations show that $\Re \left\{ \left. \partial \epsilon(q,\omega)/\partial \omega \right|_{\omega(q)} \right\}$ in (\ref{eq:time2}) becomes large when $\omega(q)$ approaches $qv$, which happens for very large wavevectors and this also reduces the probability of emission of large wavevector plasmons. Figure (\ref{fig2}) shows that the electron spontaneous emission lifetimes can range from values as small as 10 fs to hundreds of picoseconds. Figure (\ref{fig4}) shows the average minority carrier (electron) recombination time $\tau_{R}$ plotted as a function of the minority carrier density for different temperatures in p-doped graphene ($p=10^{12}$ cm$^{-2}$). As expected from the results in Figure (\ref{fig2}), the average recombination time decreases with the increase in the temperature because the minority carrier distribution spreads to higher energies. 
\begin{figure}[tbp]
  \begin{center}
   \epsfig{file=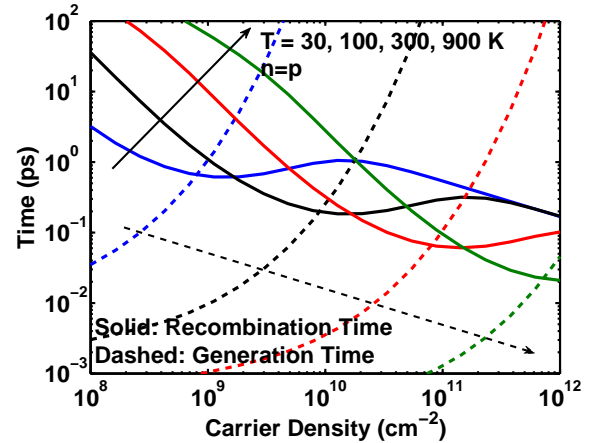,angle=-0,width=3.0 in}    
    \caption{The recombination times $\tau_{R}$ (solid) and the generation times $\tau_{G}$ (dashed) are plotted as a function of the electron and hole density (assumed to be equal) for different temperatures. The arrows indicate curves for increasing values of the temperature ($T=30,100,300,900$K).}  
    \label{fig5}
  \end{center}
\end{figure}

\begin{figure}[tp]
  \begin{center}
   \epsfig{file=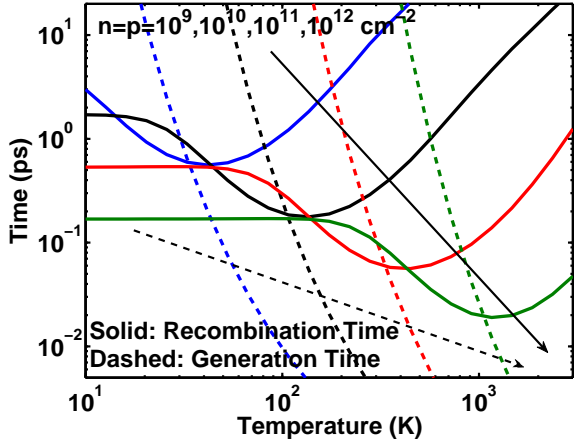,angle=-0,width=3.0 in}    
    \caption{The recombination times $\tau_{R}$ (solid) and the generation times $\tau_{G}$ (dashed) are plotted as a function of the temperature for different electron and hole densities (assumed to be equal). The arrows indicate curves for increasing values of the carrier density ($n=p=10^{9}, 10^{10}, 10^{11}, 10^{12}$ cm$^{-2}$).}  
    \label{fig6}
  \end{center}
\end{figure}

In many optical studies~\cite{rana4,rana5,heinz,heinz2,norris,driel,rana6,kaindl} and in graphene based optoelectronic devices~\cite{avouris,avouris2,mceuen,park}, photoexcitation followed by rapid thermalization results in a an equal number of thermally distributed electrons and holes in an otherwise near intrinsic graphene layer (Figure (\ref{fig0}d)). It is therefore important to understand the recombination and generation times in such situations. Figure (\ref{fig5}) shows the recombination times $\tau_{R}$ (solid) and the generation times $\tau_{G}$ (dashed) plotted as a function of the electron and hole density (assumed to be equal) for different temperatures. The number of plasmons $n(\vec{q})$ in different modes is assumed to be given by the Bose factor. This assumption may not be valid in a non-equilibrium situation immediately following photoexcitation. Figure (\ref{fig6}) shows the recombination times $\tau_{R}$ (solid) and the generation times $\tau_{G}$ (dashed) plotted as a function of the temperature for different electron and hole densities. Figures (\ref{fig5}) and (\ref{fig6}) show that the recombination times can be much smaller than a picosecond for carrier densities larger than $10^{11}$ cm$^{-2}$ at all temperatures. Figures (\ref{fig5}) and (\ref{fig6}) show that the generation times can also be very short and this implies that carrier generation cannot be ignored in experiments where a hot carrier distribution is created via photoexcitation in ultrafast optical studies~\cite{rana4,rana5,heinz,heinz2,norris,driel,rana6,kaindl}. 

\section{Theoretical Model for Graphene on Polar Substrates}
\begin{figure}[tbp]
  \begin{center}
   \epsfig{file=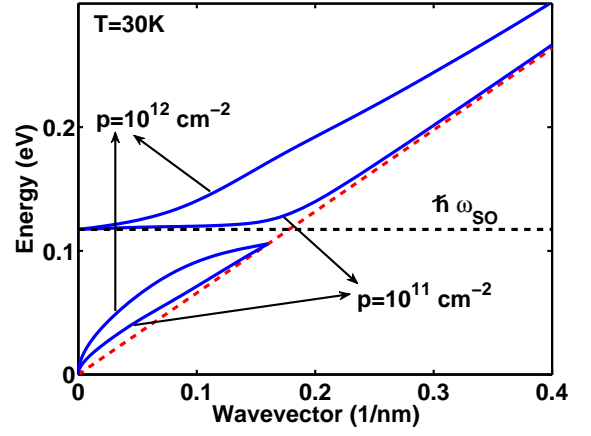,angle=-0,width=3.0 in}    
    \caption{The dispersion of the coupled plasmon-phonon mode (solid) in a p-doped graphene sheet on a SiC substrate is shown for different hole densities. The dispersion splits into two branches. The dashed curve represents $\hbar v q$. $T=30K$.}
    \label{fig7}
  \end{center}
\end{figure}

The results presented above suggest that it might be possible to alter the plasmon-assisted recombination and generation rates in graphene by altering the dielectric environment thereby modifying the strength of the Coulomb interaction~\cite{jena}. Specifically, a substrate with a large dielectric constant could potentially reduce the recombination and generation rates. However, polar materials with large dielectric constants have surface optical phonon modes that couple strongly with the graphene plasmons~\cite{koch}. To study this further, we consider a graphene sheet at a distance $d$ away from a polar substrate (Figure (\ref{fig0}b)). The dielectric constant of the substrate is assumed to be given by the expression,
\begin{equation}
\epsilon_{sub}(\omega) = \epsilon_{sub}(\infty) \left( \fracd{\omega^{2} - \omega_{LO}^{2}}{\omega^{2} - \omega_{TO}^{2}} \right)
\end{equation}
Here, $\epsilon_{sub}(0)/\epsilon_{sub}(\infty) = \omega_{LO}^{2}/\omega_{TO}^{2}$. The surface optical phonon frequency $\omega_{SO}$ is obtained by setting $\epsilon_{sub}(\omega)$ equal to -1, and equals,
\begin{equation}
 \omega_{SO} = \omega_{TO}\sqrt{\fracd{\epsilon_{sub}(0)+1}{\epsilon_{sub}(\infty)+1}}
\end{equation}
The dielectric constant $\epsilon(q,\omega)$ of the graphene sheet can be found by placing a test charge in the sheet and finding the resulting potential. The result is,
\begin{eqnarray}
\epsilon(q,\omega) & = & \fracd{1}{2} + \fracd{1}{2} \left[ \fracd{ (\epsilon_{sub}(\omega)+1) e^{2qd} + (\epsilon_{sub}(\omega)-1) }{ (\epsilon_{sub}(\omega)+1) e^{2qd} - (\epsilon_{sub}(\omega)-1) } \right] \nonumber \\
& & - \fracd{e^{2}}{2\epsilon_{o}q} \Pi(q,\omega)
\end{eqnarray}
The dispersion of the coupled plasmon-phonon longitudinal mode can be found as before by setting $\epsilon(q,\omega)$ equal to zero. Now one finds two longitudinal collective modes. In the $q\rightarrow 0$ limit, the lower frequency mode is plasmon-like with $\omega(q) \rightarrow 0$ as $q\rightarrow 0$, and the higher frequency mode is phonon-like with $\omega(q) \rightarrow \omega_{SO}$ as $q\rightarrow 0$. For large wavevectors, the lower frequency mode disappears into the electron-hole continuum while the higher frequency mode becomes plasmon-like with $\omega(q) \rightarrow qv$ as $q\rightarrow \infty$. As an example, we consider the technologically relevant case of a graphene layer on a Silicon Carbide (SiC) substrate~\cite{heer,koch,sarma}. The values of different parameters are as follows: $d=5$ Angstroms, $\hbar \omega_{LO}=120$ meV, $\hbar \omega_{TO}=98$ meV, and $\epsilon_{\infty}=6.5$~\cite{surf}. These give $\hbar \omega_{SO} \approx 117$ meV. Figure (\ref{fig7}) shows the dispersions of the coupled plasmon-phonon modes for a p-doped graphene sheet on a SiC substrate for different hole densities. Comparing Figures (\ref{fig1}) and (\ref{fig7}), it can be seen that plasmon-phonon coupling significantly modifies the dispersion and this has recently been verified experimentally~\cite{koch}.   
\begin{figure}[tbp]
  \begin{center}
   \epsfig{file=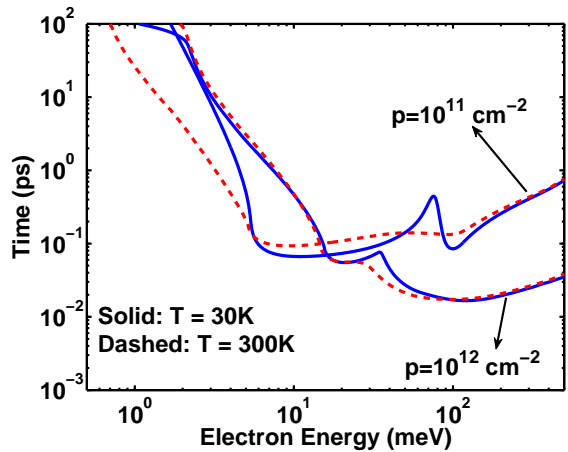,angle=-0,width=3.0 in}    
    \caption{The calculated spontaneous emission lifetime of an electron in the conduction band is plotted as a function of the electron energy for different hole densities and temperatures in p-doped graphene on a SiC substrate.}
    \label{fig8}
  \end{center}
\end{figure}
The recombination and generation rates can be obtained using the same expressions as those given in (\ref{eq:time2}), (\ref{eq:R}), and (\ref{eq:G}) with the exception that contributions from both branches of the dispersion must be included. Therefore, surface optical phonons of the polar substrate provide an additional channel for carrier recombination and generation. It should be mentioned here that large wavevector surface optical phonon modes can also cause intervalley recombination and generation processes~\cite{rana3}. However, the square of the coupling matrix element between the surface optical phonons and the carriers is proportional to~\cite{mahan2},
\begin{equation}
\fracd{e^{2}}{2 \epsilon_{o}q} e^{-2qd} \hbar \omega_{SO} \left( \fracd{1}{\epsilon_{sub}(\infty)+1} - \fracd{1}{\epsilon_{sub}(0)+1} \right)
\end{equation}
and becomes small for the large wavevectors needed for the intervalley transitions in graphene.Therefore, intervalley processes will be ignored here.   

\begin{figure}[tbp]
  \begin{center}
   \epsfig{file=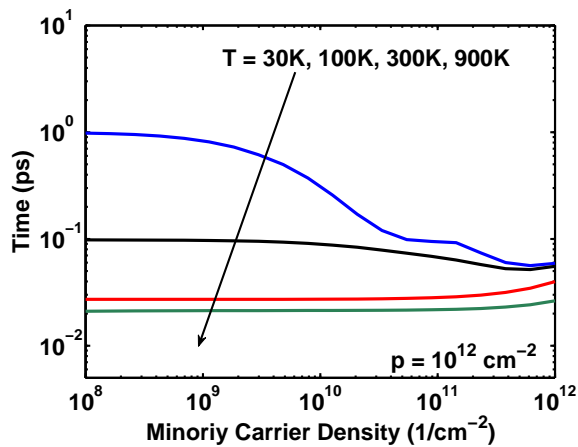,angle=-0,width=3.0 in}    
    \caption{The average minority carrier (electron) recombination time $\tau_{R}$ is plotted as a function of the minority carrier density for different temperatures in p-doped graphene ($p=10^{12}$ cm$^{-2}$) on a SiC substrate. The arrow indicates curves for increasing values of the temperature ($T=30,100,300,900$K).}  
    \label{fig9}
  \end{center}
\end{figure}
\begin{figure}[tbp]
  \begin{center}
   \epsfig{file=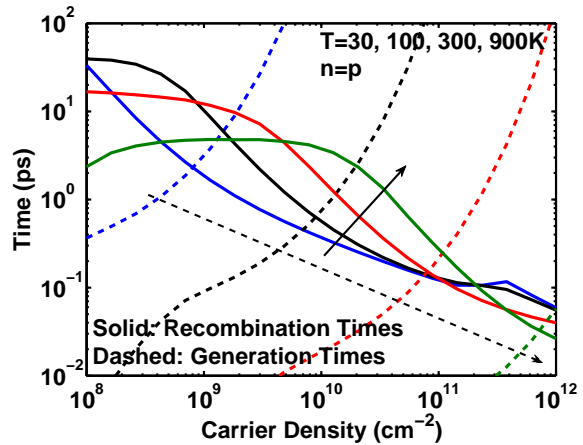,angle=-0,width=3.0 in}    
    \caption{The recombination times $\tau_{R}$ (solid) and the generation times $\tau_{G}$ (dashed) are plotted as a function of the electron and hole density (assumed to be equal) for different temperatures for a graphene sheet on a SiC substrate. The arrows indicate curves for increasing values of the temperature ($T=30,100,300,900$K).}  
    \label{fig10}
  \end{center}
\end{figure}

\subsection{Results and Discussion}
Figure (\ref{fig8}) shows the calculated lifetime of an electron in the conduction band due to spontaneous emission as a function of the electron energy for different hole densities and temperatures for a p-doped graphene sheet on a SiC substrate. Figure (\ref{fig8}) displays the same general trends as does Figure (\ref{fig2}) in the case of a suspended graphene sheet. However, lifetimes are shorter for the low energy electrons in the case of graphene on SiC. For electrons with energies near the Dirac point, recombination is entirely due to the lower frequency branch of the dispersion which facilitates interband transitions better than the plasmon dispersion in suspended graphene. The sharp peaks seen in Figure (\ref{fig8}) occur when the lifetimes due to the lower frequency branch of the dispersion are becoming longer with the electron energy while lifetimes due to the upper frequency branch are becoming shorter. As in the suspended graphene case, the spontaneous emission lifetimes can range from tens of femtoseconds to hundreds of picoseconds. Figure (\ref{fig9}) shows the minority carrier (electron) recombination time $\tau_{R}$ plotted as a function of the minority carrier density for different temperatures in a p-doped graphene ($p=10^{12}$ cm$^{-2}$) on a SiC substrate. Compared to the suspended graphene case (Figure \ref{fig4}), the recombination times for graphene on SiC are shorter for small minority carrier densities. Next, we consider the case when the electron and hole densities are the same (as is the situation in photoexcitation experiments).  Figure (\ref{fig10}) shows the recombination times $\tau_{R}$ (solid) and the generation times $\tau_{G}$ (dashed) plotted as a function of the electron and hole density (assumed to be equal) for different temperatures for a graphene sheet on a SiC substrate. The recombination and generation times in graphene on SiC are generally of the same order as in the case of suspended graphene discussed earlier. The role of the higher dielectric constant of the SiC substrate in reducing plasmon-assisted recombination and generation rates, compared to suspended graphene, is compensated by the presence of surface optical phonons which not only modify the plasmon dispersion but also provide an additional channel for recombination and generation. 

\section{Conclusion}
In this paper we have presented electron-hole recombination and generation times due to spontaneous and stimulated emission and absorption of plasmons in graphene. Our results indicate that plasmon assisted recombination times in graphene can vary over a wide range of values ranging from tens of femtoseconds to hundreds of picoseconds. In many proposed and demonstrated optoelectronic devices~\cite{rana,ryzhii,avouris,ryzhii2,ryzhii3,avouris2,ryzhii4,ryzhii5,ryzhii6,mceuen,park}, the plasmon-assisted recombination and generation rates could be fast enough to significantly impact device performance. 

\acknowledgements
The authors acknowledge helpful discussions with Paul L. McEuen, Michael G. Spencer, and Jiwoong Park, and acknowledge support from the National Science Foundation (monitor Eric Johnson), the DARPA Young Faculty Award, the MURI program of the Air Force Office of Scientific Research (monitor Harold Weinstock), the Office of Naval Research (monitor Paul Makki), and the Cornell Material Science and Engineering Center (CCMR) program of the National Science Foundation.


\newpage

\begin{thebibliography}{99}
\bibitem{nov1} K. S. Novoselov et. al., Nature, {\bf 438}, 197 (2005). 
\bibitem{nov2} K. S. Novoselov et. al., Science, {\bf 306}, 666 (2004). 
\bibitem{dressel1} R. Saito, G. Dresselhaus, M. S. Dresselhaus, {\em Physical Properties of Carbon Nanotubes}, Imperial College Press, London, UK (1999).  
\bibitem{heer} W. De Heer et. al., Science, {\bf 312}, 1191 (2006).  
\bibitem{shepard} I. Meric, M. Y. Han, A. F. Young, B. Ozyilmaz, P. Kim, K. L. Shepard, Nature Nanotechnology, {\bf 3}, 654 (2008). 
\bibitem{marcus} J. R. Williams, L. DiCarlo, C. M. Marcus, Science, {\bf 317}, 638 (2007). 
\bibitem{rana} F. Rana, IEEE Trans. Nanotech., {\bf 7}, 91 (2008). 
\bibitem{ryzhii} V. Ryzhii, M. Ryzhii, T. Otsuji, J. Appl. Phys., {\bf 101}, 083114 (2007). 
\bibitem{avouris} F. Xia, T. Mueller, Y. Lin, A. V. Garcia, P. Avouris, Nature Nanotechnology, {\bf 4}, 839 (2009). 
\bibitem{ryzhii2} F. T. Vasko,  V. Ryzhii, Phys. Rev. B, {\bf 77}, 195433 (2008).
\bibitem{ryzhii3} V. Ryzhii, M. Ryzhii, V. Mitin, T. Otsuji, J. Appl. Phy., {\bf 107}, 054512 (2010). 
\bibitem{Bon} F. Bonaccorso, Z. Sun, T. Hasan, A. C. Ferrari, Nature, {\bf 4}, 611 (2010).  
\bibitem{avouris2} T. Mueller, F. Xia, P. Avouris, Nature Photonics, {\bf 4}, 297 (2010). 
\bibitem{ryzhii4} M. Ryzhii, V. Ryzhii, T. Otsuji, V. Mitin, M. S. Shur, Phys. Rev. B, {\bf 82}, 075419 (2010). 
\bibitem{ryzhii5} V. Ryzhii, A. A. Dubinov, T. Otsuji, V. Mitin, M. S. Shur, J. Appl. Phys., {\bf 107}, 054505 (2010). 
\bibitem{ryzhii6} V. Ryzhii, M. Ryzhii, T. Otsuji, J. Appl. Phys., {\bf 101}, 083114 (2007). 
\bibitem{bostwick} A. Bostwick, F. Speck, T. Seyller, K. Horn, M. Polini, R. Asgari, A. H. MacDonald, E. Rotenberg, Science, {\bf 328}, 999 (2010). 
\bibitem{ohta} A. Bostwick, T. Ohta, T. Seyller, K. Horn, E. Rotenberg, Nature Physics, {\bf 3}, 36 (2007). 
\bibitem{polini} M. Polini, R. Asgari, G. Borghi, Y. Barlas, T. Pereg-Barnea, A. H. MacDonald, Phys. Rev. B, {\bf 77}, 081411(R) (2008).
\bibitem{sarma0} E. H. Hwang, S. Das Sarma, Phys. Rev. B, {\bf 77}, 081412(R) (2008).
\bibitem{rana2} F. Rana, Phys. Rev. B, {\bf 76}, 155431 (2007). 
\bibitem{rana3} F. Rana, P. A. George, J. H. Strait, J. Dawlaty, S. Shivaraman, M. Chandrashekhar, M. G. Spencer, Phys. Rev. B, {\bf 79}, 115447 (2009). 
\bibitem{koch} R. J. Koch, T. Seyller, J. A. Schaefer, B, {\bf 82}, 201413(R) (2010). 
\bibitem{sarma} E. H. Hwang, R. Sensarma, S. Das Sarma, Phys. Rev. B, {\bf 82}, 195406 (2010). 
\bibitem{rana4} P. A. George, J. Strait, J. Dawlaty, S. Shivaraman, Mvs Chandrashekhar, F. Rana, M. G. Spencer, Nano Lett., {\bf 8}, 4248 (2008). 
\bibitem{rana5} J. M. Dawlaty, S. Shivaraman, Mvs Chandrashekhar, F. Rana, M. G. Spencer, Appl. Phys. Lett., {\bf 92}, 042116 (2008).
\bibitem{heinz} Chun Hung Lui, Kin Fai Mak, Jie Shan, Tony F. Heinz, Phys. Rev. Lett., {\bf 105}, 127404 (2010). 
\bibitem{heinz2} Kin Fai Mak, Chun Hung Lui, Tony F. Heinz, Appl. Phys. Lett., {\bf 97}, 221904 (2010).
\bibitem{norris} D. Sun, Z. K. Wu, C. Divin, X. Li, C. Berger, W. A. de Heer, P. N. First, T. B. Norris, Phys. Rev. Lett., {\bf 101}, 157402 (2008).
\bibitem{driel} R. W. Newson, J. Dean, B. Schmidt, and H. M. van Driel, Opt. Express, {\bf 17}, 2326 (2009);
\bibitem{rana6} H. N. Wang, J. H. Strait, P. A. George, S. Shivaraman, V. B. Shields, M. Chandrashekhar, J. Hwang, F. Rana, M. G. Spencer, C. S. Ruiz-Vargas, and J. Park, Appl. Phys. Lett. 96, 081917 (2010).
\bibitem{kaindl} H. Choi, F. Borondics, D. A. Siegel, S. Y. Zhou, M. C. Martin, A. Lanzara, R. A. Kaindl, Appl. Phys. Lett., {\bf 94}, 172102 (2009).
\bibitem{bolotin} K.I. Bolotin, K.J. Sikes, Z. Jianga, M. Klima, G. Fudenberg, J. Hone, P. Kim, H.L. Stormer, Solid State Commun., {\bf 146}, 351 (2008).
\bibitem{sarma2} E. H. Hwang, S. Das Sarma, Phys. Rev. B, {\bf 75}, 205418 (2007). 
\bibitem{tapash} X. F. Wang, T. Chakraborty, Phys. Rev. B, {\bf 75}, 033408 (2007).
\bibitem{kong} D. H. Staelin, A. W. Morgenthaler, J. A. Kong, {\em Electromagnetic Waves},Prentice Hall, NJ (1998). 
\bibitem{mahan} G. D. Mahan, {\em Many Particle Physics}, Plenum Press, NY (1990). 
\bibitem{mceuen} X. Xu, N. M. Gabor, J. S. Alden, A. M. Van Der Zande, P. L. McEuen, Nano Lett., {\bf 10}, 562 (2010). 
\bibitem{park} J. Park, Y. H. Ahn, C. S. Ruiz-Vargas, Nano Lett., {\bf 9}, 1742 (2009).  
\bibitem{surf} H. Nienhaus, T. U. Kampen, W. Monch, Surf. Sci., 324, L528 (1995).
\bibitem{jena} D. Jena, A. Konar, Phys. Rev. Lett., {\bf 98}, 136805 (2007).
\bibitem{mahan2} S. Q. Wang and G. D. Mahan, Phys. Rev. B, {\bf 6}, 4517 (1972).

\end{thebibliography}
\end{document}